\documentclass[a4paper]{article}
\usepackage[latin9]{inputenc}
\usepackage{verbatim}
\usepackage{amsmath}
\usepackage{graphicx}

\makeatletter

\pdfpageheight\paperheight
\pdfpagewidth\paperwidth

\providecommand{\tabularnewline}{\\}

\usepackage{INTERSPEECH2019}

\usepackage[pdftex,pdftitle={A Real-Time Wideband Neural Vocoder at 1.6 kb/s Using LPCNet},pdfauthor={Jean-Marc Valin, Jan Skoglund}]{hyperref}

\makeatother

\begin{document}
\title{A Real-Time Wideband Neural Vocoder at 1.6 kb/s Using LPCNet}
\name{Jean-Marc Valin$^{1*}$, Jan Skoglund$^2$\thanks{$^*$This work was performed while the author was with Mozilla.}}
\address{   $^1$Amazon, Palo Alto, CA, USA\\   $^2$Google LLC, San Francisco, CA, USA}
\email{jmvalin@jmvalin.ca, jks@google.com}

\maketitle
\begin{abstract}
Neural speech synthesis algorithms are a promising new approach for
coding speech at very low bitrate. They have so far demonstrated quality
that far exceeds traditional vocoders, at the cost of very high complexity.
In this work, we present a low-bitrate neural vocoder based on the
LPCNet model. The use of linear prediction and sparse recurrent networks
makes it possible to achieve real-time operation on general-purpose
hardware. We demonstrate that LPCNet operating at 1.6~kb/s achieves
significantly higher quality than MELP and that uncompressed LPCNet
can exceed the quality of a waveform codec operating at low bitrate.
This opens the way for new codec designs based on neural synthesis
models.
\end{abstract}

\section{Introduction}

Very low bitrate parametric codecs have existed for a long time~\cite{atal1971speech,markel1974linear},
but their quality has always been severely limited. While they are
efficient at modeling the spectral envelope (vocal tract response)
of the speech using linear prediction, no such simple model exists
for the excitation. Despite some advances~\cite{griffin1985new,mccree1996,rowe1997},
modeling the excitation signal has remained a challenge. For that
reason, parametric codecs are rarely used above 3~kb/s. 

Neural speech synthesis algorithms such as Wavenet~\cite{van2016wavenet}
and SampleRNN~\cite{mehri2016samplernn} have recently made it possible
to synthesize high quality speech. They have also been used in~\cite{kleijn2018wavenet,Garbacea2019}
(WaveNet) and~\cite{klejsa2018high} (SampleRNN) to synthesize high-quality
speech from coded features, with a complexity in the order of 100~GFLOPS.
This typically makes it impossible to use those algorithms in real
time without high-end hardware (if at all). In this work, we focus
on simpler models, that can be implemented on general-purpose hardware
and mobile devices for real-time communication, and that work for
any speaker, in any language. Moreover, we target the very low bitrate
of 1.6~kb/s, which is beyond the reach of conventional waveform speech
coders. 

To reduce computational complexity, WaveRNN~\cite{kalchbrenner2018efficient}
uses a sparse recurrent neural network (RNN). Other models use linear
prediction to remove the burden of spectral envelope modeling from
the neural synthesis network~\cite{lpcnet,juvela2018speaker,wang2019}.
That includes our previous LPCNet work~\cite{lpcnet} (summarized
in Section~\ref{sec:WaveRNN_and_LPCNet}), which augments WaveRNN
with linear prediction to achieve low complexity speaker-independent
speech synthesis.

We now address quantization of the LPCNet features to achieve low-bitrate
speech coding (Section~\ref{sec:Features_and_quantization}). Section~\ref{sec:Training}
discusses training considerations, and Section~\ref{sec:Evaluation}
presents our results. We conclude with ideas for improvement in Section~\ref{sec:Conclusion}.

\section{LPCNet Overview}

\label{sec:WaveRNN_and_LPCNet}

\begin{figure}
\centering{\includegraphics[width=1\columnwidth]{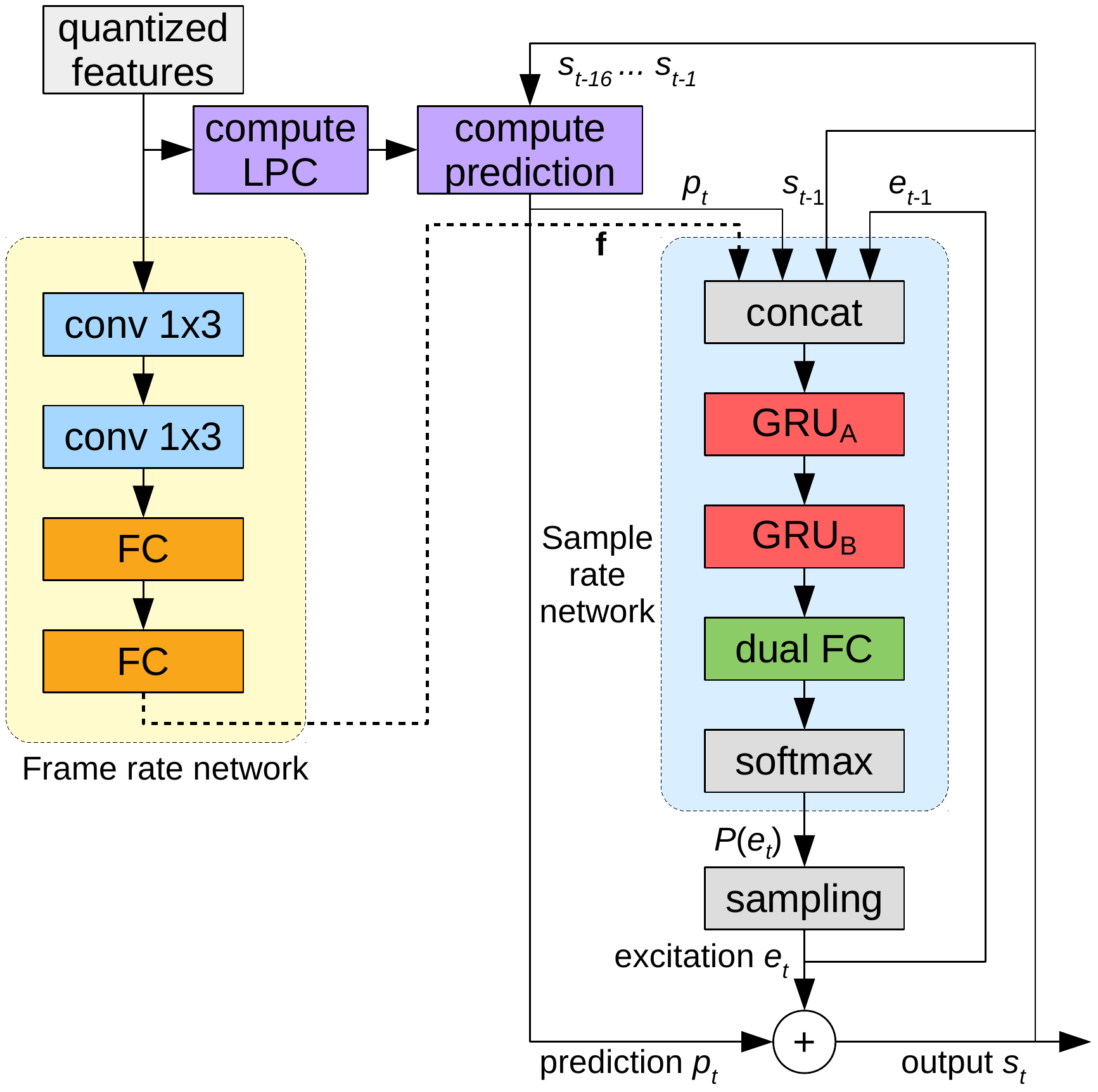}}

\caption{Overview of the LPCNet model. The frame rate network (yellow) operates
on 10\protect\nobreakdash-ms frames and its output is held constant
through each frame for the sample rate network (blue). The \emph{compute
prediction} block applies linear prediction to predict the sample
at time $t$ from the previous samples. Conversions between $\mu$-law
and linear are omitted for clarity. The de-emphasis filter is applied
to the output $s_{t}$.\label{fig:Overview-of-LPCNet}}
\end{figure}

The WaveRNN model~\cite{kalchbrenner2018efficient} is based on a
sparse gated recurrent unit (GRU)~\cite{cho2014properties} layer.
At time $t$, it uses the previous audio sample $s_{t-1}$, as well
as frame conditioning parameters to generates a discrete probability
distribution $P\left(s_{t}\right)$ from which the output $s_{t}$
is sampled. LPCNet~\cite{lpcnet} improves on WaveRNN by adding linear
prediction, as shown in Fig.~\ref{fig:Overview-of-LPCNet}. It is
divided in two parts: a frame rate network that computes conditioning
features for each 10\nobreakdash-ms frame, and a sample rate network
that computes conditional sample probabilities. In addition to using
the previously generated speech sample $s_{t-1}$, LPCNet also uses
the $16^{th}$ order prediction $p_{t}=\sum_{i=1}^{16}a_{i}s_{t-i}$
and the previously generated excitation $e_{t-1}$, where $e_{t}=s_{t}-p_{t}$. 

LPCNet operates on signals quantized using 256-level $\mu$\nobreakdash-law.
To avoid audible quantization noise we apply a pre-emphasis filter
$E(z)=1-\alpha z^{-1}$ on the input speech (with $\alpha=0.85$)
and the inverse (de-emphasis) filter on the output. This shapes the
noise and makes it less perceptible. Considering that $s_{t-1}$,
$p_{t}$, and $e_{t-1}$ are discrete, we can pre-compute the contribution
$\mathbf{v}_{i}^{\left(\cdot,\cdot\right)}$ of each possible value
to the $\mathrm{GRU_{A}}$ gates and state so that only lookups are
necessary at run-time. In addition, the contribution $\mathbf{g}^{\left(\cdot\right)}$
of the frame rate network to $\mathrm{GRU_{A}}$ can be computed only
once per frame. After these simplifications, only the recurrent matrices
$\mathbf{W}_{\left(\cdot\right)}$ remain and the sample rate network
is computed as
\begin{align}
\mathbf{u}_{t}= & \sigma\left(\mathbf{W}_{u}\mathbf{h}_{t-1}+\mathbf{v}_{s_{t-1}}^{\left(u,s\right)}+\mathbf{v}_{p_{t}}^{\left(u,p\right)}+\mathbf{v}_{e_{t-1}}^{\left(u,e\right)}+\mathbf{g}^{\left(u\right)}\right)\nonumber \\
\mathbf{r}_{t}= & \sigma\left(\mathbf{W}_{r}\mathbf{h}_{t-1}+\mathbf{v}_{s_{t-1}}^{\left(r,s\right)}+\mathbf{v}_{p_{t}}^{\left(r,p\right)}+\mathbf{v}_{e_{t-1}}^{\left(r,e\right)}+\mathbf{g}^{\left(r\right)}\right)\label{eq:LPCNet}\\
\widetilde{\mathbf{h}}_{t}= & \tau\left(\mathbf{r}_{t}\circ\left(\mathbf{W}_{h}\mathbf{h}_{t-1}\right)+\mathbf{v}_{s_{t-1}}^{\left(h,s\right)}+\mathbf{v}_{p_{t}}^{\left(h,p\right)}+\mathbf{v}_{e_{t-1}}^{\left(h,e\right)}+\mathbf{g}^{\left(h\right)}\right)\nonumber \\
\mathbf{h}_{t}= & \mathbf{u}_{t}\circ\mathbf{h}_{t-1}+\left(1-\mathbf{u}_{t}\right)\circ\widetilde{\mathbf{h}}_{t}\nonumber \\
P\left(e_{t}\right) & =\mathrm{softmax}\left(\mathrm{dual\_fc}\left(\mathrm{GRU_{B}}\left(\mathbf{h}_{t}\right)\right)\right)\,,\nonumber 
\end{align}
where $\sigma\left(x\right)$ is the sigmoid function, $\tau\left(x\right)$
is the hyperbolic tangent, $\circ$ denotes an element-wise vector
multiply, and $\mathrm{GRU_{B}}\left(\cdot\right)$ is a regular (non-sparse)
GRU. The dual fully-connected (dual\_fc) layer is defined as
\begin{equation}
\mathrm{dual\_fc}(\mathbf{x})=\mathbf{a}_{1}\circ\tau\left(\mathbf{W}_{1}\mathbf{x}\right)+\mathbf{a}_{2}\circ\tau\left(\mathbf{W}_{2}\mathbf{x}\right)\,,\label{eq:dual_fc}
\end{equation}
where $\mathbf{W}_{1}$ and $\mathbf{W}_{2}$ are weight matrices
and $\mathbf{a}_{1}$ and $\mathbf{a}_{2}$ are scaling vectors.

Throughout this paper, biases are omitted for clarity. The synthesized
excitation sample $e_{t}$ is obtained by sampling from the probability
distribution $P\left(e_{t}\right)$ after lowering the \emph{temperature}
of voiced frames as described in eq.~(7) of \cite{lpcnet}. As a
way of reducing the complexity, $\mathrm{GRU_{A}}$ uses sparse recurrent
matrices with non-zero blocks of size 16x1 to ensure efficient vectorization.
Because the hidden state update is more important than the reset and
update gates, we keep 20\% of the weights in $\mathbf{W}_{h}$, but
only 5\% of those in $\mathbf{W}_{r}$ and $\mathbf{W}_{u}$, for
an average of 10\%. 

\section{Features and Quantization}

\label{sec:Features_and_quantization}

LPCNet is designed to operate with 10\nobreakdash-ms frames. Each
frame includes 18~cepstral coefficients, a pitch period (between
16 and 256~samples), and a pitch correlation (between 0 and~1).
To achieve a low bitrate, we use 40\nobreakdash-ms packets, each
representing 4~LPCNet frames. Each packet is coded with 64~bits,
allocated as shown in Table~\ref{tab:Bit-allocation}, for a total
bitrate of 1.6~kb/s (constant bitrate). The prediction coefficients
are computed using the quantized cepstrum.

In addition to the packet size of 40~ms, LPCNet has a synthesis look-ahead
of 25~ms: 20~ms for the two 1x3 convolutional layers in the frame
rate network and 5~ms for the overlap in the analysis window (see
Section~\ref{subsec:cepstrum}). The total algorithmic delay of the
codec is thus 65~ms. Because the complexity of the frame-level processing
is negligible compared to that of the sample rate network, no significant
delay is added by the computation time (unlike traditional codecs
where up to one frame delay can be added if the codec takes 100\%~CPU).

\subsection{Pitch}

Extracting the correct pitch (without period doubling or halving)
is very important for a vocoder since no residual is coded to make
up for prediction errors. During development, we have observed that
unlike traditional vocoders, LPCNet has some ability to compensate
for incorrect pitch values, but only up to a point. Moreover, that
ability is reduced when the cepstrum is quantized. 

The pitch search operates on the excitation signal. Maximizing the
correlation over an entire 40\nobreakdash-ms packet does not produce
good results because the pitch can vary within that time. Instead,
we divide each packet in 5\nobreakdash-ms sub-frames, and find the
set of pitch lags $\tau_{i}$ that maximize
\begin{equation}
J=\sum_{i}\left[w_{i}r\left(\tau_{i}\right)-\Theta\left(\tau_{i}-\tau_{i-1}\right)\right]\,,\label{eq:pitch_cost}
\end{equation}
where $w_{i}$ is the ratio of the sub-frame $i$ energy over the
average energy of the 40\nobreakdash-ms packet, $\Theta\left(\Delta\tau\right)$
is a transition penalty defined as 
\begin{equation}
\Theta\left(\Delta\tau\right)=\begin{cases}
0.02\left(\Delta\tau\right)^{2} & \mathrm{if}\,\left|\Delta\tau\right|\leq4\\
6 & \mathrm{otherwise}
\end{cases}\,,\label{eq:pitch_transition}
\end{equation}
and $r\left(\tau\right)$ is a modified pitch correlation~\cite{vos2013voice}
\begin{equation}
r\left(\tau\right)=\frac{2\sum_{n}e\left(n\right)e\left(n-\tau\right)}{\sum_{n}e^{2}\left(n\right)+\sum_{n}e^{2}\left(n-\tau\right)}\,.\label{eq:pitch_correlation}
\end{equation}

The optimal path can be computed efficiently using dynamic programming
with a Viterbi search. Since the entire audio is not available at
once, the values of $J$ in the forward pass are updated normally
for every sub-frame but the Viterbi backtrack pass is computed once
per 40\nobreakdash-ms packet, over all 8~sub-frames. While that
does not guarantee finding the global optimal path, it ensures a consistent
pitch over the duration of the packet, which is important for quantization.

The pitch is allowed to vary between 62.5~Hz and 500~Hz. The average
pitch over the packet is encoded on a logarithmic scale using 6~bits,
resulting in quantization intervals of 0.57~semitones. 

A linear pitch modulation parameter allows up to 16\% variation (2.5~semitones)
between the first and last sub-frame. It is encoded with 3 bits representing
a $\left[-3,3\right]$ range with two different codes for 0 (constant
modulation), one of which signaling that the pitch correlation is
less than 0.3 (the modulation is zero when the pitch correlation is
too small). Two extra bits refine the value of the pitch correlation
within the remaining range (either $\left[0,\,0.3\right]$ or $\left[0.3,\,1\right]$).

\begin{table}
\caption{Bit allocation for a 40\protect\nobreakdash-ms frame.\label{tab:Bit-allocation}}
\centering{%
\begin{tabular}{lc}
Parameter & Bits\tabularnewline
\hline 
Pitch period & 6\tabularnewline
Pitch modulation & 3\tabularnewline
Pitch correlation & 2\tabularnewline
Energy (C0) & 7\tabularnewline
Cepstrum VQ (40~ms) & 30\tabularnewline
Cepstrum delta (20~ms) & 13\tabularnewline
Cepstrum interpolation (10~ms) & 3\tabularnewline
\hline 
Total & 64\tabularnewline
\end{tabular}}
\end{table}

\subsection{Cepstrum}

\label{subsec:cepstrum}

\begin{figure}
\centering{\includegraphics[width=0.9\columnwidth]{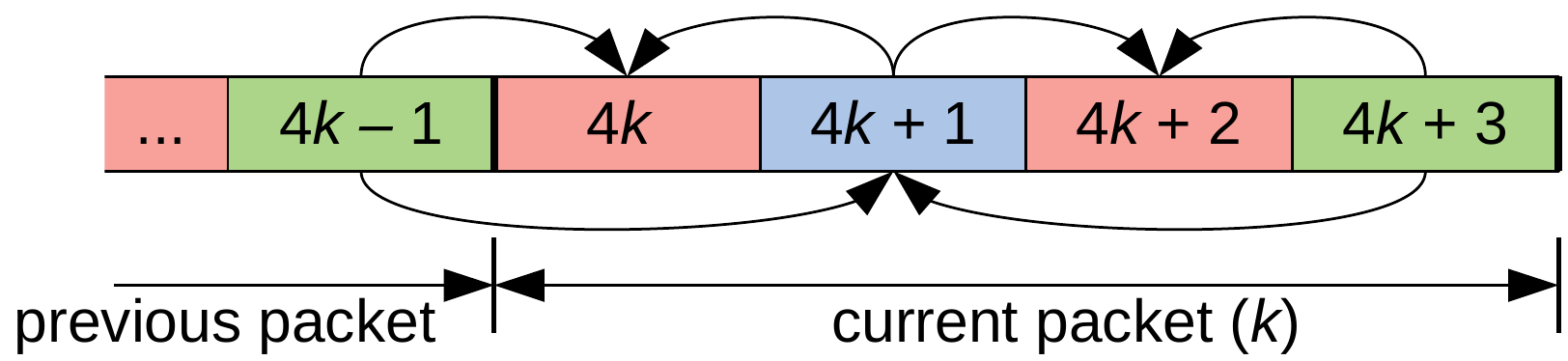}}

\caption{Prediction and quantization of the cepstrum for packet~$k$. Vectors
in green are quantized independently, vectors in blue are quantized
with prediction, and vectors in red use prediction with no residual
quantization. Prediction is shown by the arrows. \label{fig:Prediction-and-quantization}}
\end{figure}

The spectral feature analysis operates on 20\nobreakdash-ms windows
with a 10\nobreakdash-ms frame offset (50\% overlap). The cepstrum
is computed from 18~Bark-spaced bands following the same layout as~\cite{valin2017hybrid}.
Because we pack 4~cepstral vectors in each packet, we wish to minimize
redundancy within a packet while limiting dependencies across packets
to reduce the effect of packet loss. For those reasons, we use a prediction
scheme (Fig.~\ref{fig:Prediction-and-quantization}) inspired by
video codec B\nobreakdash-frames~\cite{MPEG1}, limiting the error
propagation in case of packet loss to a worst case of 30~ms.

Let packet $k$ include cepstral vectors $\mathbf{c}_{4k}$ to $\mathbf{c}_{4k+3}$.
We start by coding the first component (C0) of $\mathbf{c}_{4k+3}$
independently, with a uniform 7\nobreakdash-bit scalar quantizer
(0.83~dB resolution). We then code the remaining component of $\mathbf{c}_{4k+3}$
independently using a 3\nobreakdash-stage codebook with 17~dimensions
and 10~bits for each stage. An M-best (survivor) search helps reduce
the quantization error, but is not strictly necessary. From there,
vector $\mathbf{c}_{4k+1}$ is predictively coded using both $\mathbf{c}_{4k-1}$
(independently coded in the previous packet) and $\mathbf{c}_{4k+3}$.
We use a single bit to signal if the prediction is the average ($\frac{\mathbf{c}_{4k-1}+\mathbf{c}_{4k+3}}{2}$),
or two bits if the prediction is either of $\mathbf{c}_{4k-1}$ or
$\mathbf{c}_{4k+3}$. The 18\nobreakdash-dimensional prediction residual
is then coded with a 11\nobreakdash-bit + sign codebook for the average
predictor or with an 10\nobreakdash-bit + sign codebook if not, for
a total of 13~bits for $\mathbf{c}_{4k+1}$. Although the average
predictor is the most useful, including the single-vector predictors
improves the quantization of transients/onsets. 

Because there is insufficient bitrate to adequately code $\mathbf{c}_{4k}$
and $\mathbf{c}_{4k+2}$, we only use a prediction from their neighbors.
Vector $\mathbf{c}_{4k}$ is predicted from its neighbors $\mathbf{c}_{4k-1}$
and $\mathbf{c}_{4k+1}$, whereas $\mathbf{c}_{4k+2}$ is predicted
from $\mathbf{c}_{4k+1}$ and $\mathbf{c}_{4k+3}$. Since there are
3~options for each vector, we have 9~possible combinations. By eliminating
the least useful combination (forcing both equal to $\mathbf{c}_{4k+1}$),
we can code the remaining ones with 3~bits.

\section{Training}

\label{sec:Training}

\begin{figure}
\centering{\includegraphics[width=1\columnwidth]{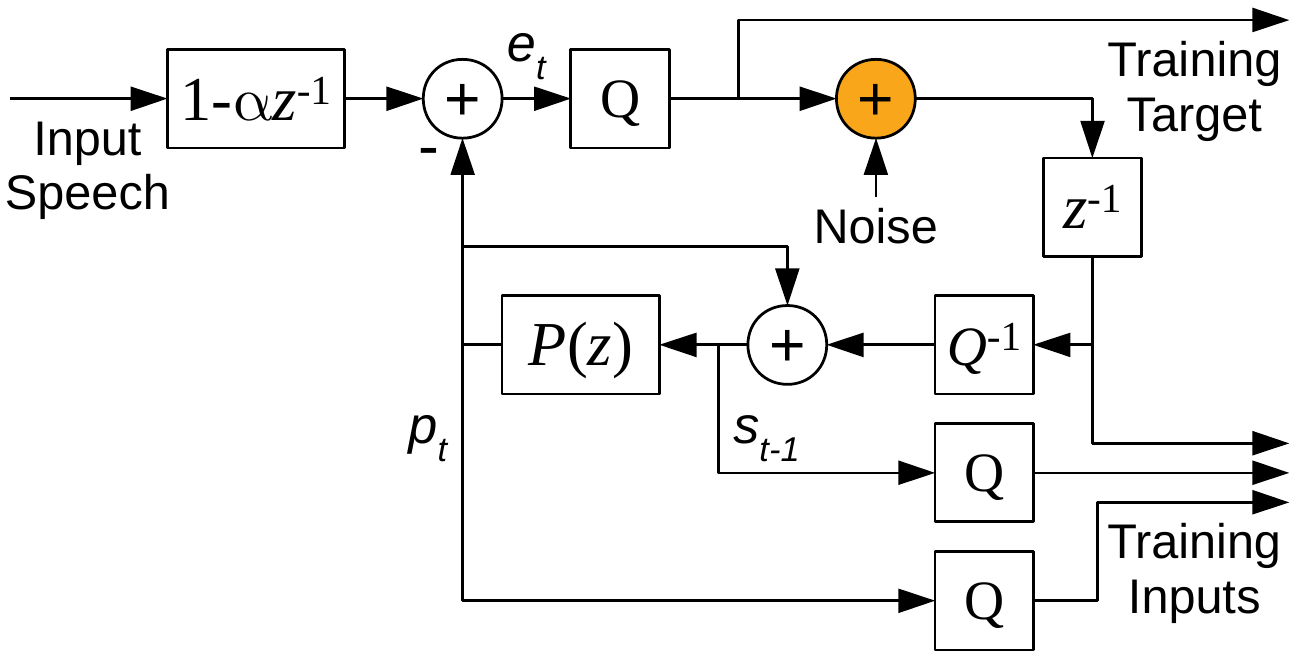}}

\caption{Noise injection during the training procedure, with $Q$ denoting
$\mu$-law quantization and $Q^{-1}$ denoting conversion from $\mu$-law
to linear. The prediction filter filter is given by $P\left(z\right)=\sum_{i=1}^{M}a_{i}z^{-k}$.
The target excitation is computed as the difference between the clean,
unquantized input and the noisy prediction. Note that the noise is
added in the $\mu$-law domain so that its power follows that of the
real excitation signal.\label{fig:Noise-injection}}
\end{figure}

When decoding speech, the sample rate network operates on the synthesized
speech rather than on the \emph{ground-truth} clean speech. That mismatch
between training and inference can increase the speech distortion,
in turn increasing the mismatch. To make the network more robust to
the mismatch, we add noise to the training data, as originally suggested
in~\cite{jin2018fftnet}. As an improvement over our previous work~\cite{lpcnet},
Laplace-distributed noise is added to the excitation inside the prediction
loop, making the signal and excitation noise consistent with each
other, as shown in Fig.~\ref{fig:Noise-injection}.

As a way of making the model more robust to input variations, we use
data augmentation on the training database. The speech level is varied
over a 40~dB range and the frequency response is varied according
to eq.~(7) in~\cite{valin2017hybrid}. 

Training is first performed with unquantized features. When training
a model for quantized features, we start with a model trained on unquantized
features, and apply domain adaptation to quantized features. We have
observed that better results are obtained when only the frame rate
network is adapted, with the sample rate network weights left unchanged.
In addition to the slightly better quality, this has the advantage
of faster training for new quantizers and also smaller storage if
different quantized models are needed.

\section{Evaluation}

\label{sec:Evaluation}

The source code for this work is available under a BSD license at
\url{https://github.com/mozilla/LPCNet/}. The evaluation in this
section is based on commit \texttt{6fda6b7}.

\subsection{Complexity and Implementation}

The number of weights in the sample rate network is approximately
\begin{equation}
W=3dN_{A}^{2}+3N_{B}\left(N_{A}+N_{B}\right)+2N_{B}Q\,,\label{eq:weight_count}
\end{equation}
where $N_{A}$ and $N_{B}=16$ are the sizes of the two GRUs, $d$~is
the density of the sparse GRU, and $Q=256$ is the number of $\mu$-law
levels. Based on the subjective results in~\cite{lpcnet}, we consider
$N_{A}=384$ with $d=10\%$ (122~dense equivalent units) to provide
a good compromise between quality and complexity. This results in
$W=71600\,\mathrm{weights}$, which fits in the L2 or L3 cache of
most modern CPUs. Considering that each weight is used once per sample
for a multiply-add, the resulting complexity is 2.3~GFLOPS. The activation
functions are based on a vectorized exponential approximation and
contribute 0.6~GFLOPS to the complexity, for a total of 3~GFLOPS
when counting the remaining operations.

A C implementation of the decoder (with AVX2/FMA intrinsics) requires
20\% of a 2.4~GHz Intel Broadwell core for real-time decoding (5x
faster than real-time). According to our analysis, the main performance
bottleneck is the L2 cache bandwidth required for the matrix-vector
products. On ARMv8 (with Neon intrinsics), real-time decoding on a
2.5~GHz Snapdragon~845 (Google Pixel~3) requires 68\% of one core
(1.47x~real-time). On the more recent 2.84~GHz Snapdragon~855 (Samsung
Galaxy~S10), real-time decoding requires only 31\% of one core (3.2x~real-time).

As a comparison, we estimate the complexity of the SampleRNN neural
codec described in~\cite{klejsa2018high} to be around 100~GFLOPS
-- mostly from the MLP with two hiddens layers and 1024 units per
layer. The complexity of the WaveNet-based codec in~\cite{kleijn2018wavenet}
significantly exceeds 100~GFLOPS\footnote{based on discussion with the authors}.
For speaker-dependent text-to-speech~(TTS) -- which typically allows
smaller models -- real-time synthesis was achieved by~\cite{arik2017deep_rt}
(WaveNet) and~\cite{kalchbrenner2018efficient} (WaveRNN) when using
multiple CPU cores.

The total complexity of the proposed encoder is around 30~MFLOPS
(0.03~GFLOPS), mostly from the 5-best VQ search (14~MFLOPS) and
the undecimated 16~kHz pitch search (8~MFLOPS). Although the encoder
complexity could be significantly reduced, it is already only 1\%
of the decoder complexity.

\subsection{Experimental Setup}

The model is trained using 4~hours of speech from the NTT Multi-Lingual
Speech Database for Telephonometry (21~languages), from which we
excluded all samples from the speakers used in testing. From the original
data, we generate 14~hours of augmented speech data as described
in Section~\ref{sec:Training}. The unquantized network was trained
for 120~epochs (625k~updates), with a batch size size of 64, each
sequence consisting of 2400~samples (15~frames). Training was performed
on an Nvidia GPU with Keras\footnote{\url{https://keras.io/}}/Tensorflow\footnote{\url{https://www.tensorflow.org/}}
using the CuDNN GRU implementation and the AMSGrad~\cite{reddi2018convergence}
optimization method (Adam variant) with a step size $\alpha=\frac{\alpha_{0}}{1+\delta\cdot b}$
where $\alpha_{0}=0.001$, $\delta=5\times10^{-5}$, and $b$ is the
batch number. For model adaptation with quantized features, we used
40~epochs (208k~updates) with $\alpha_{0}=0.0001$, $\delta=0$.

\subsection{Quality}

\begin{figure}
\centering{\includegraphics[width=1\columnwidth]{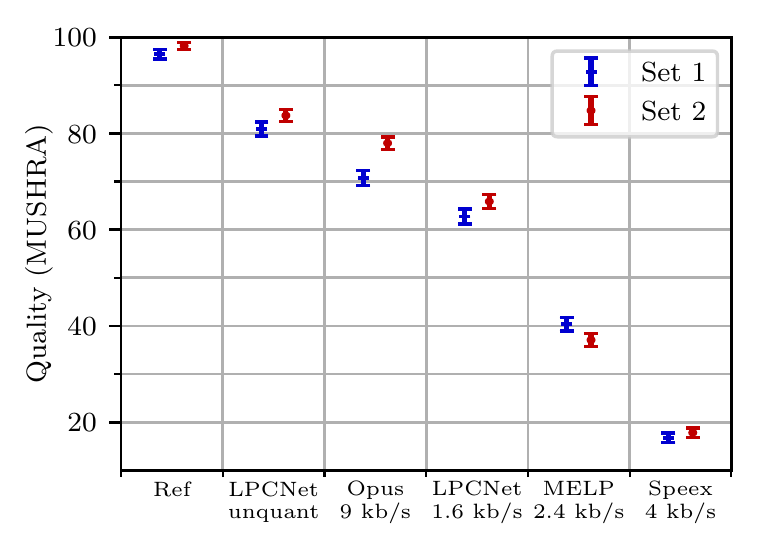}}

\caption{Subjective quality (MUSHRA) results for both listening tests. Set~1
is taken from the NTT database, while Set 2 consists of Opus testvector
samples.\label{fig:Subjective-quality-MUSHRA}}
\end{figure}

We conducted a subjective listening test with a MUSHRA-inspired methodology~\cite{BS1534}
to evaluate the quality of the proposed 1.6~kb/s neural vocoder.
As an upper bound on the quality achievable with LPCNet at the target
complexity (higher quality is achievable with a larger model), we
include LPCNet operating on unquantized features. We also compare
with Opus~\cite{rfc6716} wideband (SILK mode) operating at 9~kb/s
VBR\footnote{the lowest bitrate for which the encoder defaults to wideband}
and with the narrowband MELP~\cite{mccree1996} vocoder. As low anchor,
we use Speex~\cite{valin2007speex} operating as a 4~kb/s wideband
vocoder (wideband quality~0).

In a first test (Set~1), we used 8~samples from 2~male and 2~female
speakers. The samples are part of the NTT database used for training,
but all samples from the selected speakers for the test were excluded
from the training set. As reported in~\cite{klejsa2018high}, mismatches
between the training and testing database can cause a significant
difference in the output quality. We measure that impact in a second
test (Set~2) on the same model, with 8~samples (one~male and one~female
speaker) from the sample set used to create the Opus testvectors.
Each test included 100~listeners. 

The results in Fig.~\ref{fig:Subjective-quality-MUSHRA} show that
1.6~kb/s LPCNet clearly outperforms the MELP vocoder, making it a
viable choice for very low bitrates. The fact that LPCNet with unquantized
features achieves slightly higher quality than Opus at 9~kb/s suggests
that LPCNet at bitrates around 2-6~kb/s may be able to compete with
waveform coders below 10~kb/s. The test with samples outside the
NTT database shows that the LPCNet model generalizes to other recording
conditions. 

Since the LPCNet model was trained on 21~languages, it is expected
to also work in those languages. While we only tested on English,
informal listening indicates that the quality obtained on French,
Spanish, and Swedish is comparable to that on English.

A subset of the samples from the listening test is available at {\footnotesize\url{https://people.xiph.org/~jm/demo/lpcnet_codec/}}. 

\section{Conclusions}

\label{sec:Conclusion}

We have demonstrated a 1.6~kb/s neural vocoder based on the LPCNet
model that can be used for real-time communication on a mobile device.
The quality obtained exceeds what is achievable with existing low-bitrate
vocoders such as MELP. Although other work has demonstrated similar
or higher speech quality at similar bitrates~\cite{kleijn2018wavenet,Garbacea2019,klejsa2018high},
we believe this is the first neural vocoder that can operate in real-time
on general-purpose hardware and mobile devices. We have so far focused
on clean, non-reverberant speech. More work is needed for testing
and improving the robustness to noise and reverberation. 

Considering that uncompressed LPCNet is able to achieve higher quality
than 9~kb/s Opus, we believe it is worth exploring higher LPCNet
bitrates in the 2\nobreakdash-6~kb/s range. Moreover, in the case
of a waveform codec like Opus operating at very low bitrate (\textless ~8kb/s)
it should be possible to directly use the encoded LSPs to synthesize
a higher quality output than the standard provides. In addition, using
the encoded excitation as an extra input to LPCNet may help further
reducing artifacts, turning LPCNet into a neural post-filter that
can significantly improve the quality of any existing low-bitrate
waveform codec.

\section{Acknowledgements}

We thank David Rowe for his feedback and suggestions, as well as for
implementing some of the Neon optimizations.

\bibliographystyle{IEEEtran}
\bibliography{lpcnet}

\end{document}